# Deep sub-micron normally-off AlGaN/GaN MOSFET on silicon with $V_{TH}$ > 5V and On-Current > 0.5 A/mm

Sandeep Kumar, Sandeep Vura, Surani B. Dolmanan, Sudhiranjan Tripathy, R. Muralidharan, Digbijoy N. Nath *Member* IEEE

*Abstract*— We report on the demonstration of a deep sub-micron normally-off AlGaN/GaN HEMT with high on-current and high threshold voltage ($V_{TH}$). The high-performance device was realized by utilizing a gate recess with length and depth of 200 nm and 124 nm respectively. The recess etched region had a roughness of 0.7 nm. Various recess etch depths and dielectric annealing conditions were used to tune the $V_{TH}$. The optimized device exhibited on-current and $V_{TH}$ of 500 mA/mm and 5 V respectively. The measured breakdown characteristics of the devices and their limitations were investigated using 2D-TCAD device simulation. The penetration of the residual electric field in most of the recess region could be the reason for premature breakdown of deeply scaled recess-gate e-mode HEMTs

*Index Terms*—2-dimensional electron gas (2DEG), GaN, Normally-off, e-mode, Recess etched, High Electron Mobility Transistor (HEMT).

## I. INTRODUCTION

III-NITRIDE power devices are becoming increasingly popular due to their superior figures of merit compared to state-of-art of silicon (Si) power MOSFETs [1].

Table. 1 Comparison of start of the art normally off device

| $V_{TH}$ (V) from Linear extrapolation | $I_D$ (mA/mm) | Reference (Device Type) |
|---|---|---|
| +7.6 | 355 | [3] Recess MOSFET |
| +3.6 | 430 | [4] (F⁻ implanted HEMT) |
| +5.2 | 200 | [5] (Recess MOSFET) |
| +7.2 | 120 | [6] (Recess MOSFET) |
| +1.0 | 200 | [7] (p-AlGaN) |
| +3.0 | 260 | [8] (p-GaN) |
| +4.3 | 27 | [9] (p-GaN) |
| +6.5 | 340 | [10] (Partially etched barrier +F⁻ implanted gate dielectric) |
| **+5.1** | **500** | **This Work** |

We acknowledge funding support from MHRD through NIEIN project, from MeitY and DST through NNetRA.
S. Kumar, S. Vura, R. Muralidharan, and Digbijoy N. Nath are with Centre for Nano Science and Engineering (CeNSE), Indian Institute of Science (IISc), Bengaluru, India (e-mail: sandeepku@iisc.ac.in)
S. B. Dolmanan and S. Tripathy are with Institute of Materials Research and Engineering (IMRE), Agency for Science, Technology, and Research (A*STAR), Singapore

GaN HEMTs are normally on by virtue of the growth stack; however, several power electronics applications demand normally-off devices for safety concerns [2]. Several device architectures/ technologies have been explored for the realization of normally-off devices; however, the pursuit for high on-current and high $V_{TH}$ devices is still continuing. Most of the reported normally-off devices suffer from either low on-current or low $V_{TH}$ (Table 1). Here, we report on a sub 500 nm recess etched normally-off HEMT device with high on-current and high $V_{TH}$.

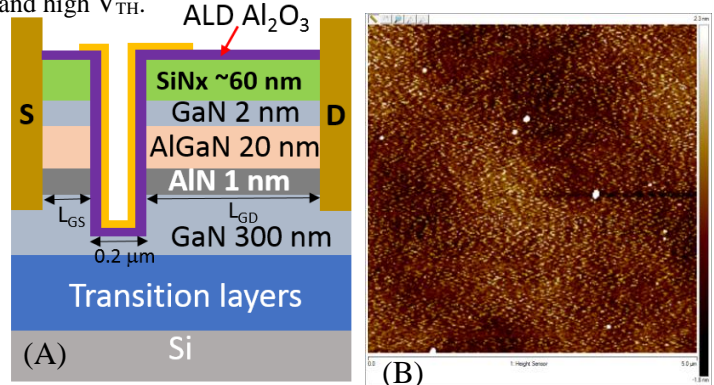

Fig. 1 (A) Epitaxial stack and device architecture (B) AFM image of recess region.

## II. EPI-STACK DETAILS AND DEVICE FABRICATION

The epi-stack (Fig. 1(A)) growth details can be found in ref [11]. Device fabrication started with e-beam evaporation of Ti/Al/Ni/Au for ohmic contacts and alignments marks (Optical and e-beam lithography). Ohmic contacts were rapid thermal annealed at 850° C for 30 sec in $N_2$ ambience. Mesa etching was done using Cl-based reactive ion etching and mesa etch depth was of ~300 nm. 60 nm of PECVD SiNx was deposited for recess etch mask. 200 nm of recess length was opened using e-beam lithography. SiNx was first etched in the recess region using $CHF_3$ and $O_2$ RIE chemistry and then barrier and some thickness of GaN channel were etched using $Ar/BCl_3/Cl_2$ chemistry. The roughness of the etched region was measured to be 0.7 nm from AFM (Fig 1(B)). Before atomic layer deposition (ALD) $Al_2O_3$ deposition, sample was cleaned with $NH_4OH$, HCl and HF. 30 nm of $Al_2O_3$ was deposited using ALD for gate dielectric. The recess depth and post dielectric anneal temperatures were varied to achieve a high $V_{TH}$ and high on-current device. Post dielectric anneal (PDA) at 700 °C and 500 °C were done in forming gas ambience for 1 min. Ni/Au metal was e-beam evaporated for Schottky gate and the samples were placed at an angle of 10 °C from the horizontal for better sidewall coverage. Post gate metal anneal (PMA) was done for 5 min in FGA at 500 °C. Finally, $Al_2O_3$ and SiNx above the source-drain pads were dry etched for probing the device. Three types of devices, device I (recess etch depth 60 nm (barrier+channel), PDA 700°C 1 min, PMA 500°C 5 min), device II (recess etch depth 40 nm (barrier+channel), PDA 500°C 1 min, PMA 500°C 5 min) and device III (recess etch depth 124 nm (barrier+channel), PDA 500°C 1 min, PMA 500°C 5 min) are discussed in this paper. The SEM image of fabricated device is shown in Fig 2.

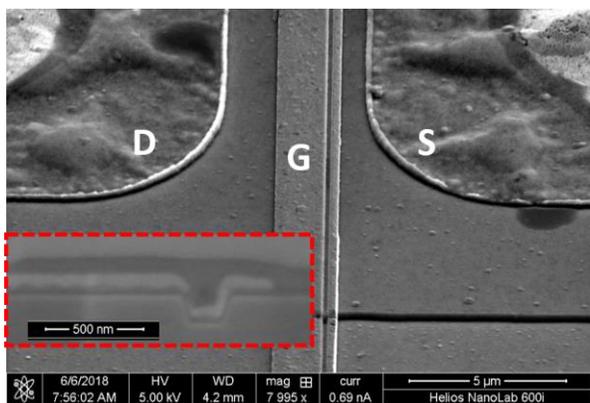

Fig. 2. SEM image of fabricated device (Device III), cross-section of gate region is shown as inset (in red dashed box)

### III. DEVICE CHARACTERIZATION AND ANALYSIS

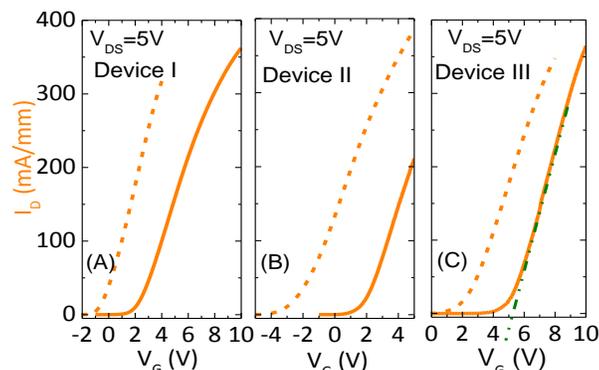

Fig. 3. Transfer characteristics of device (A) $L_{gs}$=1 μm, $L_g$=200 nm, $L_{fp}$=2.6 μm, $L_{gd}$=9 μm, (B) $L_{gs}$=1 μm, $L_g$=200 nm, $L_{fp}$=2.5 μm, $L_{gd}$= 9 μm, (C) $L_{gs}$=1.2 μm, $L_g$=200 nm, $L_{fp}$=3 μm, $L_{gd}$=11.4 μm.

The transfer and output characteristics of the devices are shown in Fig. 3 and Fig. 4 respectively. The device I (Fig. 3 (A)) exhibited a $V_{TH}$ of ~-1V before PMA and the $V_{TH}$ was found to shift to ~2V after PMA. The increase in the $V_{TH}$ should be related to the decrease in the fix charge at the interface of $Al_2O_3$/GaN as reported previously [12][13][14]. The recess depth of the device was 60 nm (barrier + channel) as measured by AFM. As the high temperature PDA had resulted in a negative $V_{TH}$ of the device (before PMA), a lower PDA temperature was implemented for subsequent device processes.

The transfer characteristics of device II is shown in Fig. 3(B). The recess depth in this case was 40 nm (barrier + channel). The device II exhibited a $V_{TH}$ of -3V and 2V before PMA and after PMA. The gate leakage was also found to reduce from 1 μA/mm to 0.1 nA/mm after gate metal annealing (not shown). As the $V_{TH}$ was found to be 2V even with shallower etch depths with lower PDA temperatures, a deep recess etching was implemented to further increase the $V_{TH}$.

Fig. 3 (C) shows the transfer characteristics of device III. The recess depth in this case was 124 nm (barrier + channel). The $V_{TH}$ was found to be 2 V and >5 V before PMA and after PMA. The drain current was > 350 mA/mm at gate and drain bias of 10V and 5V.

The output characteristics of fabricated devices are shown in the Fig. 4. The devices exhibited similar output characteristics except the output conductance worsened from (A) to (C). The device III exhibited excellent on-current (500 mA/mm) and $V_{TH}$ (>5V) combinations reported till date to the best of our knowledge. The devices were further characterized for 3-terminal breakdown characteristics.

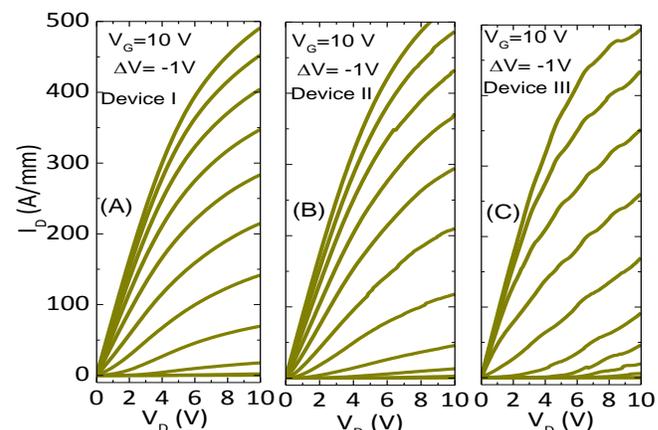

Fig. 4 Output characteristics of devices I, II, and III.

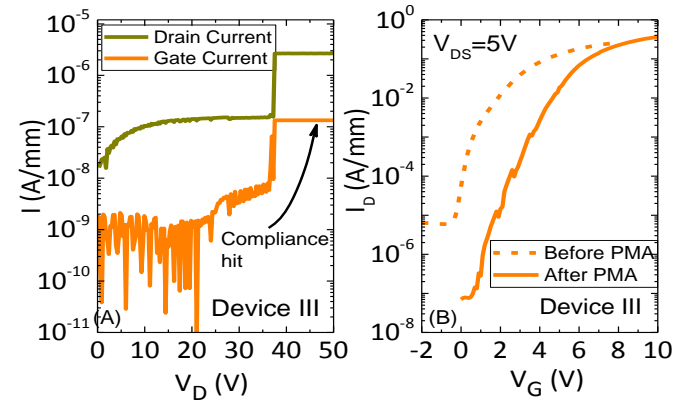

Fig. 5. (A) Three terminal breakdown characteristics and (B) transfer characteristics of device III.

The three terminal breakdown characteristics of device III (deeply recessed) are shown in Fig. 5 (A). The device III was found to break destructively ~ 40 V. Devices I and II exhibited similar breakdown characteristics. The subpar breakdown characteristics could be attributed to the oxide/nitride interface and it can be inferred from the transfer characteristics in log scale (Fig. 5 (B)). The subthreshold drain current was found to be rising slowly with gate voltage (<$V_{TH}$) and further improvements are required on this front.

The low breakdown characteristics were further investigated by Silvaco TCAD simulations of short and long channel recess devices. The material parameters used in the device simulation were defined under kp.set2 and pol.set2 in Silvaco ATLAS. The HEMT stack considered in simulation had 24 nm of $Al_{0.25}Ga_{0.75}N$ barrier on 300 nm UID GaN ($10^{15}$ cm$^{-3}$) channel. The GaN buffer was 600 nm. The gate lengths were 200 nm and 2 μm for short and long gate recessed devices respectively while access regions were fixed at 500 nm (source side) and 4 μm (drain side). 30 nm of $Al_2O_3$ was used as gate dielectric. The recess depth was 35 nm (AlGaN barrier=25 nm + GaN channel 10 nm).

The simulated electric field profiles below the gate along the drain for short and long gate length devices are shown in the Fig. 6 (A) and Fig. 6(B). The residual electric field was found to be penetrating most of the recess region of shorter recess length device while the residual electric field had penetrated ~200 nm into the recess region for longer recess length device. The poor breakdown characteristics of 200 nm recess device could be related to field induced carrier transport through the $Al_2O_3$/etched GaN interface. Current leakage through the interfaces are widely reported in III-Nitrides [15][16].

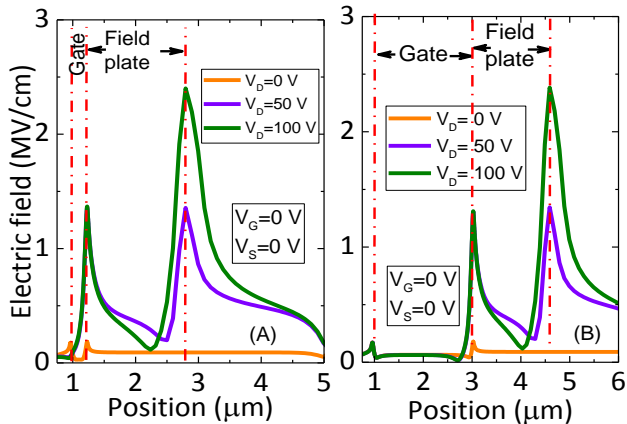

Fig. 6 (A) Electric field profile below the gate for (A) short gate device and (B) long gate device.

## IV. CONCLUSIONS

A recess etched normally-off AlGaN/GaN HEMT with excellent on current and $V_{TH}$ was demonstrated. Various recess depths and annealing conditions were investigated and optimized. The subpar breakdown characteristics were investigated using Silvaco TCAD simulations and it could be attributed to the penetration of residuals electric fields in most of the recess region. Further enhancement in the breakdown characteristics can be achieved by a slightly larger recess length and/or improved $Al_2O_3$/etched GaN interface.

## ACKNOWLEDGMENT

This publication is an outcome of the Research and Development work undertaken in the Project under Ph.D. scheme of Media Lab Asia. Authors would like to acknowledge the National NanoFabrication Centre (NNFC) and Micro and Nano Characterization Facility (MNCF) at CeNSE, IISc for device fabrication and characterization.